# Astro2020 Science White Paper

# Precision Analysis of Evolved Stars

**Thematic Areas:** Stars and Stellar Evolution

**Principal Author:**
Name:        Stephen Ridgway
Institution:   NOAO
Email:         ridgway@noao.edu
Phone:        520-248-4408

**Co-authors:** Rachel Akeson (Caltech/IPA), Ellyn Baines (NRO), Michelle Creech-Eakman (NMT/MRO), Tabetha Boyajian (LSU), Elvire De Beck (ChalmersUT), Andrea Dupree (CfA), Doug Gies (GSU), Kenneth Hinkle (NOAO), Elizabeth Humphreys (ESO), Roberta Humphreys (UMinn), Richard Joyce (NOAO), Lynn Matthews (MIT), John Monnier (U MIch), Ryan Norris (GSU), Rachael Roettenbacher (Yale), Letizia Stanghellini (NOAO), Theo ten Brumellaar (GSU), Gerard van Belle (Lowell), Wouter Vlemmings (ChalmersUT), J Craig Wheeler (UT Austin), Russell White (GSU), Lucy Ziurys (UA)

**Abstract:** Evolved stars dominate galactic spectra, enrich the galactic medium, expand to change their planetary systems, eject winds of a complex nature, produce spectacular nebulae and illuminate them, and transfer material between binary companions. While doing this, they fill the HR diagram with diagnostic loops that write the story of late stellar evolution. Evolved stars sometimes release unfathomable amounts of energy in neutrinos, light, kinetic flow, and gravitational waves. During these late-life times, stars evolve complexly, with expansion, convection, mixing, pulsation, mass loss. Some processes have virtually no spatial symmetries, and are poorly addressed with low-resolution measurements and analysis. Even a "simple" question as how to model mass loss resists solution. However, new methods offer increasingly diagnostic tools. Astrometry reveals populations and groupings. Pulsations/oscillations support study of stellar interiors. Optical/radio interferometry enable 2-3d imagery of atmospheres and shells. Bright stars with rich molecular spectra and  velocity fields are a ripe opportunity for imaging with high spatial and spectral resolution, giving insight into the physics and modeling of later stellar evolution.

# The Ubiquity of Evolved Stars in Astrophysics

Evolution of abundances in stars is the result of nucleosynthesis, but evolution of elements in galaxies is the product of dynamic processes that distribute these products. The spectra of evolved stars and of inter/circumstellar matter trace galactic abundances in our neighborhood, and to the beginning of observable time. Creating a defensible history of elemental evolution from singularity to today will rely on sufficient understanding of nearby stars to extrapolate that knowledge to early times and forward.

## Element Synthesis and Beyond

The story of the lighter elements post-big bang begins in the vicinity of the main sequence, and continues on the giant branch. Stellar evolution theory developed a half-century ago, with on-going laboratory support, has now gained observational leverage with the study of solar-like oscillations, giving precious access to structure of single stars, providing essential parameters such as absolute radii, and such remarkable information as identification of the dominant energy producing reactions (Bedding et al., 2011).

Low mass stars mostly lock up their nuclear products forever. Stars that evolve to the giant branch release a rich variety of materials in one or more stages of mass ejection. This includes a variety of isotopes of common elements, which can be used to constrain nucleosynthetic processes. The low density, low pressure, low heat capacity, but opaque atmospheres of giant stars can support variously oscillations and convection. Depending on temperature, gravity and luminosity, stellar atmospheres support levitation of surface material by mass motion, shock waves, and radiation pressure on dust grains, molecules or ions, producing ejection of substantial fractions of the stellar mass, including synthesized materials. The process is impacted by the interconversions between physical states of the ejected material. The details of mass loss are under study. Approximate scaling relations are generally assumed. Convection and oscillation (probably with overtone modes) must interact. Companions are a wild card. Factors such as magnetic fields and rotation are only suspected. There is a long way to go in this science.

Mass loss produces obvious signatures such as planetary nebulae and classical novae, and obscure such as thick dust envelopes. Depending on the stellar mass and mass loss, some evolve to white dwarfs, candidates for SN. An important subset produces core collapse supernovae, with bulk ejection and further (poorly known) nuclear processing. Resulting black holes and neutron stars participate in interacting binaries, and are candidates for high energy interactions that produce detectable gravitational waves. These complex stars will never give up all their secrets. Fortunately, understanding the broad sweep of galaxy evolution, and the major contents of the Milky Way, does not require dissecting every bizarre stellar config-uration. It behooves us to make use of the tools available to study the important and complex processes that dominate the story. Advanced stellar evolution is one of the most essential.

The thesis of this paper is that techniques for the study of late stellar evolution are well-known, ripe for vigorous exploitation and further development. Here is a fast, richly illustrated tour. Additional details will be provided in other papers.

## Opportunities for study

**Synoptic Photometry Surveys.** A tidal wave of photometric surveys (SDSS, Pan-STARRS, ZTF, GAIA, LSST, …) with rapidly improving data quality and curation will carry forward the study of stellar evolution with direct benefits. Databases will support exhaustive empirical "classification" with large numbers of even rare objects. Long time-baselines will discover targets in short-lived phases, through secular changes in brightness, color, and periodicity.

**Fast Photometry Surveys.** The Corot and Kepler missions produced a reconnaissance archive that continues to drive development of the field of asteroseismology. Harmonic analysis of time-series provides insight into the interior structure of stars from the main sequence, up the giant branch, and even to white dwarfs. Core nuclear phase, depth of the convection zone, internal rotation, even inclination of the rotation axis can be backed out of time-series data. PLATO (2026 launch) will extend this opportunity to thousands of stars around the sky, supporting statistical extension of the fruits of seismic interiors investigations to targets too faint for direct measurements.

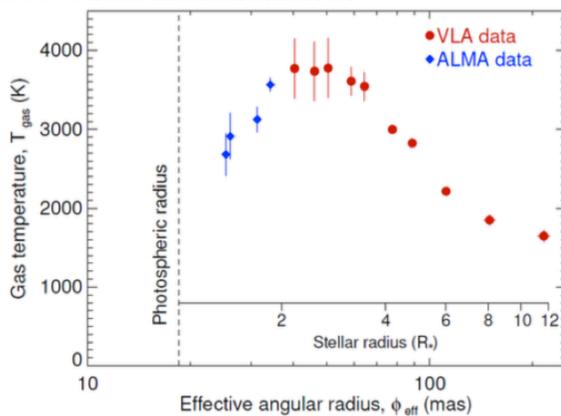

Temperature structure of the MOLsphere, chromosphere, wind transition regions of Antares from radio continuum tomography (O'Gorman, E., 2018)

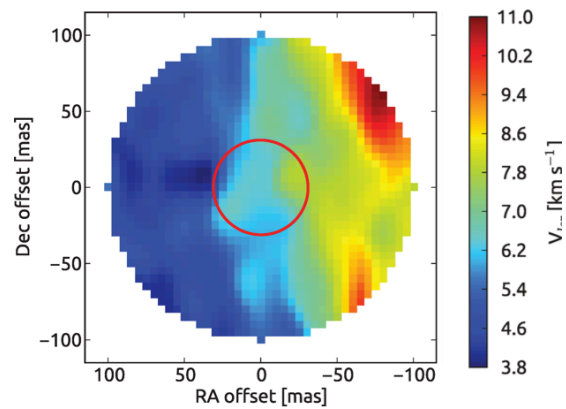

Apparent "solid body" rotation in the R Dor extended atmosphere from $SO_2$ – (circle representing 214 GHz photosphere), Vlemmings et al., 2018

**Spectroscopic Surveys.** The growing onslaught of massively multiplexed spectroscopic surveys (SDSS-V, HETDEX, APOGEE, LAMOST, GAIA, 4MOST, DESI, MSE …), will contribute to a growing reservoir of homogeneous spectroscopic datasets. Multiplicity and abundances are zero'th order parameters for understanding stars, and spectroscopy can lift the degeneracies that complicate understanding of large photometric samples. Molecular-line surveys in the millimeter of AGB, post-AGB, and even PNe can reveal chemical and elemental compositions, and produce carbon, oxygen, nitrogen, silicon, and other isotope ratios that are important tracers of interior nucleosynthesis.

**Astrometry.** With GAIA, for the first-time astronomers will have an astrometric dataset of sufficient coverage and precision that it can serve as a go-to resource for stellar astronomy and physics. The essential task of converting fluxes and angular measures to luminosity and linear measure is achieved with a combination of distance and radiometric calibrations.

**Adaptive optics imaging.** The drive to build AO imaging systems optimized for high contrast to detect exoplanets is producing a steady return in high quality imagery of circumstellar environments, with resolution down to the stellar surface for bright GB and AGB stars. Polarization maps are particularly diagnostic.

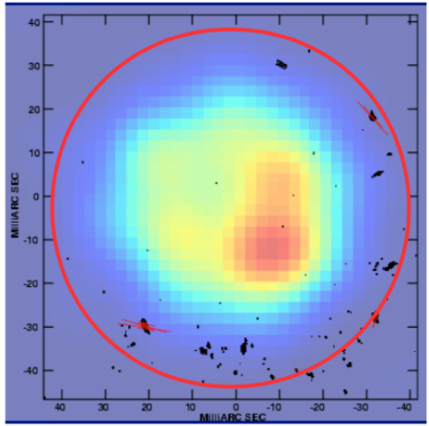 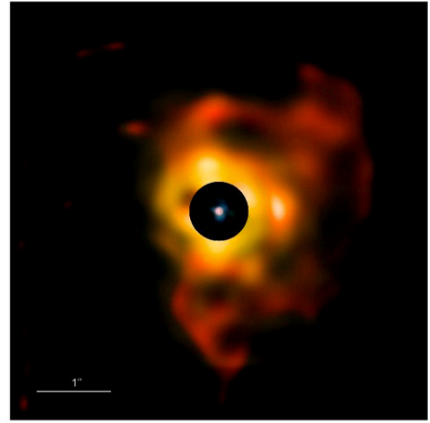

W Hya radio 338 GHz photosphere (color, Vlemmings et al. 2017); 22 GHz photosphere (circle); SiO masers (contours) – Richards et al.,2018

Circumstellar environment of Betelgeuse, coded (graphic color – wavelength in µm) blue-1.2, green-1.6, red-2.2, yellow-8.59, orange-10.48, red-17.65 (Kervella et al., 2011)

**Spectroscopic/polarametric Imaging**. Imaging methods improve steadily. Space-based UV methods (HST, BRITE) are proving particularly powerful. Temporal studies of both hot and cool luminous stars show in 2-D the dynamic heterogeneity of gas motions in chromospheres, complementing advances in interferometric imaging of molecular gas and dust in mass loss.

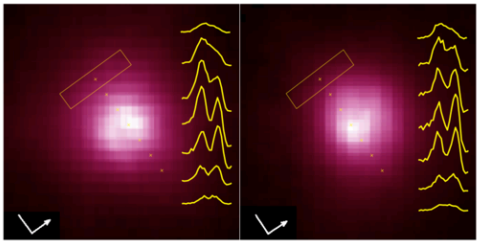 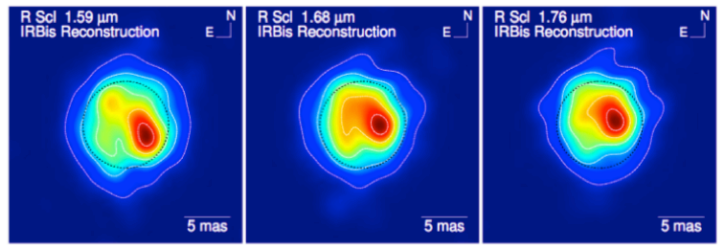

Spatially resolved UV (Si I) spectroscopy of the chromosphere of Betelgeuse, showing time evolution over 5 months (Lobel & Dupree, 2001). Slit length 100 marcsec.

Wavelength dependent appearance of presumed convection cell structure in R Scl (Wittkowski et al., 2017)

**Spectro-Interferometry**. Interferometric resolution of stars has advanced dramatically since the last decadal, with CHARA/VLTI/NPOI in the optical, and ALMA in the millimeter. These methods are powerful, with much potential in the accessible near-future. Imaging reveals the structural and temperature distributions associated with rotation, pulsation, convection, and mass loss, from the MS to the GB, AGB and post-ABG. Interferometry exploits spectroscopy, and delivers velocity-indexed data cubes, to probe complex molecular regimes (stellar atmospheres and shells) in 3-dimensional detail. Evolved stars are especially suited for these methods, and bright, prototype objects, are nearly ideal laboratories, opening stellar physics on their convection scales to time-domain study. Spatial resolution of Zeeman-broadening is within sight. The expansion of galactic novae can now be measured from the first day. The capabilities needed to link evolution of circumstellar and photospheric structures is mature and awaits systematic multi-technique exploitation. AGB and post-ABG, and the planetary nebula phase.

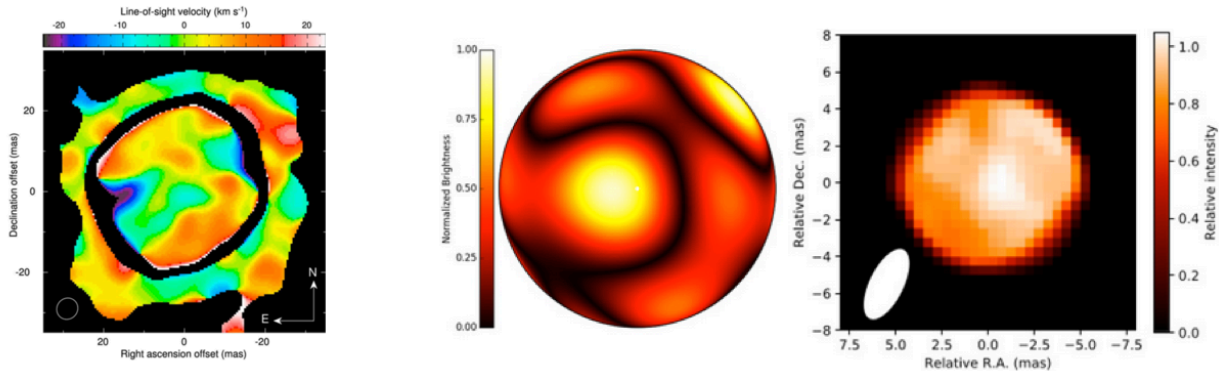

Line-of-sight atmospheric motion in Antares from near-IR CO (Ohnaka et al., 2017)

CE Tau ( left) polarization at 588 nm (Lopez et al., 2018); (right) nearly coeval H-band image (Montarges et al., 2018)

**Advanced Modeling.** Classical modeling will gain renewed vigor with the arrival of large databases representing different populations, formation environments, ages and abundances. For more complex modeling, hydrodynamic codes attained the algorithmic sophistication to address evolved stars more than a decade ago. Models are in use for tomographic analysis of stellar spectrointerferometry. Massively parallel processing makes it possible now to run those codes for experimentation, and soon for systematic investigation.

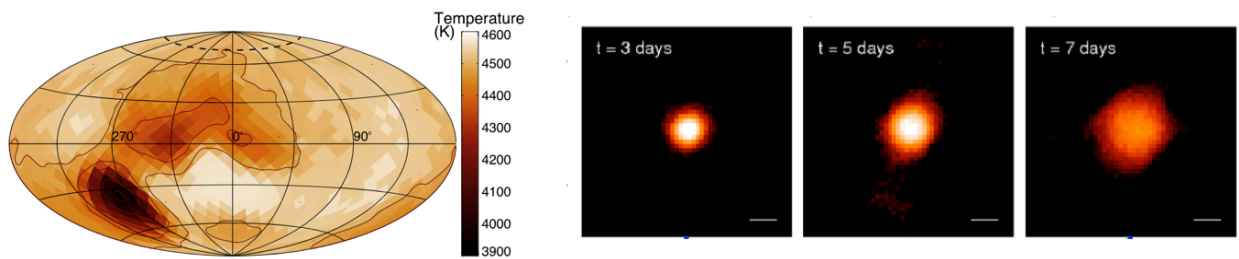

Spots on the RSCvn star sigma Gem in the H band (Roettenbacher, et al. ,2017)

Nova Delphini in H-band, at t = 3, 5 and 7 days, length of bar 1 marcsec (Schaefer, et al., 2014).

**Supporting Laboratory Science.** The rich elemental and molecular spectra of evolved stars contribute to the complexity of the physics and of the data. The relevant energy levels, transition frequencies and probabilities, and complete and accurate sets of spectroscopic constants are of only slight interest to the atomic/molecular physics community – the motivation for collecting essential data for relatively "simple" but often spectroscopically-complex species such as observed in stars requires the support of our community.

## Foreseeable Impacts

The revolution in stellar physics will transpire over the next decades as new generations of scientists grow up familiar with the capabilities and methods currently in early exploitation.

The optimum uses of massive data archives, time-series information, and multi-spectral interferometric measurements requires a learning process, with an associated period of dedicated applications and interpretation. As usual with stellar physics, the stories will be pieced together by association of objects in differing states of evolution, linked with increasingly comprehensive modeling. It is realistic to look forward to determining with good probability the ages, compositions, and interior states of bright stars, to extend this understanding statistically to large numbers of galactic stars, and further, to the interpretation of integrated spectra from distant galaxies and the oldest observable stellar light.

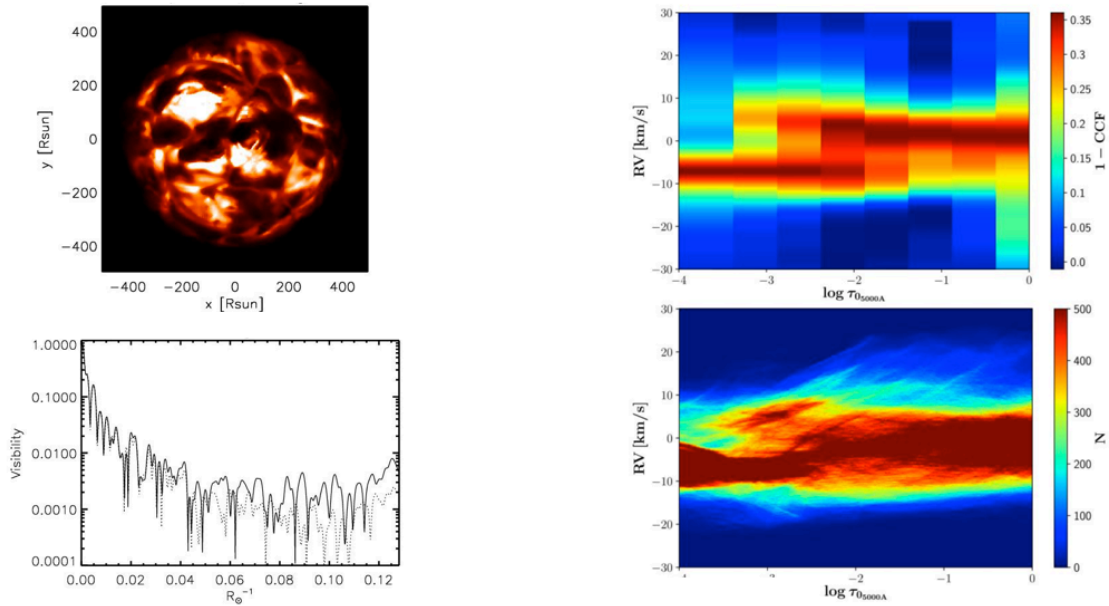

Above - image in 720 nm TiO band from 3-d hydrodynamic/radiative simulations of red supergiant atmosphere; Below – associated interferometric visibility (Chiavassa et al., 2011).

Supergiant velocity structure as a function of optical depth and radial velocity by tomographic imaging, from model, with simulated interferometric measurements – below, model; above, result of analysis (Kravchenko et al., 2018)

## Recommendations

Synoptic photometry – temporal extension of on-going/coming surveys

Spectroscopic surveys – massively multiplexed R=5-10 and 20K all-sky stellar survey

Astrometry – Proper motion improvement extension to GAIA; data curation

Adaptive optics imaging – bright object precision AO with ELTs

Spectroscopic imaging – optical/IR integral field spectroscopy for ELTs

Optical and radio interferometric imaging – 3-10X angular resolution gains

Spectro-interferometry – interferometry instrumentation/detector development

Advanced modeling – 10x increase in hardware parallelism

Laboratory science – Baseline on-going support for atomic and molecular physics

Basic support - archiving & data curation, research grants, student/early career opportunities